# Analysis of Fields in an Air-cored Superconducting Synchronous Motor with an HTS Racetrack Field Winding

Di Hu, Jin Zou, Tim J. Flack, Xiaozhuo Xu, Haichao Feng, and Mark D. Ainslie, *Member, IEEE*

*Abstract*--High temperature superconducting (HTS) synchronous motors can offer significant weight and size reductions, as well as improved efficiency, over conventional copper-wound machines due to the higher current density and lower resistance of HTS materials. In order to optimize the design parameters and performance of such a machine, this paper proposes a basic physical model of an air-cored HTS synchronous motor with a copper armature winding and an HTS field winding. An analytical method for the field analysis in the synchronous motor is presented, followed by two-dimensional (2-D) and three-dimensional (3-D) numerical finite element analysis (FEA) models to verify the analytical solution. The theoretical solution is utilized to study the influence of the geometry of the HTS coils on the magnetic field at the armature winding, and geometrical parameter optimization is carried out to obtain a more sinusoidal magnetic field at the armature winding, which has a major influence on the performance of the motor.

## I. INTRODUCTION

IN recent years, the high temperature superconducting (HTS) synchronous motor for ship propulsion has drawn a great deal of attention for its expected higher power density, higher efficiency, and smaller volume [1]-[4]. As a result, there are numerous companies and institutions who have devoted themselves to the construction of such HTS machines, such as American Superconductor Corporation (AMSC) [5], General Electrical Company (GE) [6] and Siemens [7].

D. Hu and J. Zou would like to acknowledge financial support from Churchill College, the China Scholarship Council and the Cambridge Commonwealth, European and International Trust. M. D. Ainslie would like to acknowledge financial support from a Royal Academy of Engineering Research Fellowship. This work is partly supported by a Henan International Cooperation Grant: 144300510014.

D. Hu, J. Zou and M. D. Ainslie are with the Bulk Superconductivity Group, Department of Engineering, University of Cambridge, Cambridge CB2 1PZ, UK (e-mail: dh455@cam.ac.uk, jz351@cam.ac.uk, mda36@cam.ac.uk).

T. J. Flack is with the Electrical Engineering Division, Department of Engineering, University of Cambridge, CAPE Building, Cambridge CB3 0FY, UK (e-mail: tjf1000@cam.ac.uk).

X. Xu and H. Feng are with the School of Electrical Engineering and Automation, Henan Polytechnic University, Jiaozuo, Henan 454000, P.R. China (e-mail: xxz@hpu.edu.cn, fhc@hpu.edu.cn)

Because of the high magnetic field in the HTS machine, an air core is preferred to avoid iron saturation, hysteresis and excessive heating [8]. The machine design replaces the conventional iron core with an air core design and thus, has a lighter weight and smaller size. HTS materials can carry high DC current without resistance [9]; therefore, HTS coils are frequently used to replace copper to produce high magnetic field as field winding and conventional copper windings are used on the stator, such as the 5MW HTS motor made by AMSC and ALSTOM [5].

Before fabrication, most institutions use the finite element method to design the machine and simulate its performance [10], [11]. However, few researchers have considered theoretical and analytical methods for the machine's magnetic field analysis.

It is important to develop a theoretical model for such an air-cored HTS synchronous motor with a copper armature winding and racetrack HTS coils as the DC field winding. The theoretical model can help understand the physical mechanism of the machine and help design and optimize the machine with small amount of computing time. Therefore, in this paper, a general model of an HTS synchronous motor with an HTS field winding is proposed here for field analysis.

The paper is arranged as follows. In Section II, the simplified structure of an HTS synchronous motor is presented. In Section III, the analytical model for the HTS synchronous motor is proposed, including the current distribution in the windings, the field analysis of the synchronous motor, and the reactance calculation for the motor. In Section IV, two-dimensional (2-D) and three-dimensional (3-D) numerical models of the HTS motor have been developed to compare with the analytical solution. In Section V, the advantages of the theoretical model are presented. In Section VI, the potential applications of the theoretical results are proposed, including optimization of the coil geometry by analyzing the magnetic field at the armature winding. A Kim-like model [12] is employed to verify that the coils are operated within the limit of the critical current, and the further optimization of magnetic fields over 1 T at the armature winding is completed.

## II. SIMPLIFIED STRUCTURE OF HTS SYNCHRONOUS MOTOR

A simplified structure of an HTS synchronous motor is proposed with the most important components, including the rotor, stator, air gap, armature windings and field windings. In Fig.1, $r_b$ is the radius of the rotor, $r_{s1}$ is the inner radius of the stator and $r_{s2}$ is the outer radius of the stator. The armature winding is located at $r_{s1}$ and the field winding is located at $r_{s2}$. As the thickness of the windings is much smaller than that of motor, current sheets can be used to represent the current in the windings. The armature windings are conventional copper windings, which are represented simply by sinusoidal current sheets. The field windings are HTS racetrack coils, which are distributed as current sheets at $r_b$.

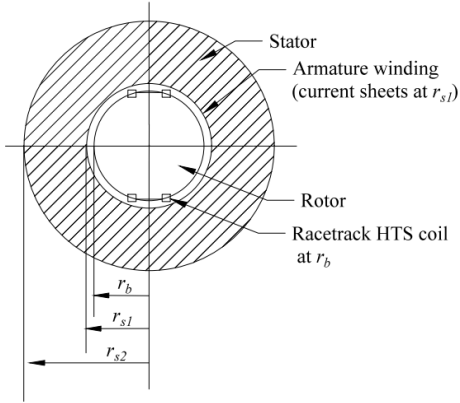

Fig. 1. The simplified structure of the HTS synchronous motor with racetrack HTS coils

In the analysis, assuming the field is 2-dimensional and the model's length is much larger than its diameter, end effects are neglected in this model. Therefore, all the current in the windings flows only perpendicular to the page (i.e., in and out of the page). For the fully air-cored synchronous machine, the permeability $\mu = \mu_0$ in all regions, where $\mu_0$ is the permeability of free space. The specific analysis for current sheets in the windings and field analysis is shown in Section III.

## III. ANALYTICAL CALCULATION OF MOTOR PARAMETERS

### A. Winding Current Distribution

For a $p$ pole-pair motor, considering the number of turns in series per phase ($N_s$), number of phases ($m$), $n^{th}$ space harmonic winding factors ($k_{wn}$), diameters ($D_{s1} = 2\ r_{s1}$), the magnitude of the stator current in one phase ($I_{ph}$), the angular frequency ($\omega$), the individual phase current of an $m$-phase a.c. winding is

$$i_k(t) = I_{ph} \cos(\omega t - \frac{2\pi}{m}(k-1)), k = 1, 2,..m \quad (1)$$

Set the space angle $\theta = 0$ on the longitudinal axis and complete the Fourier analysis of the $m$-phase currents in the space and sum the contributions for m phases. Considering the specific time t = 0, the equivalent current sheet $K_A(\theta)$ [13] for a conventional armature winding is shown as

$$K_A(\theta) = \sum_{n=2mz+1} K_{An} \sin(np\theta), z = 0,1,2... \quad (2)$$

$$K_{An} = \frac{2mN_s I_s k_{wn}}{\pi D_{s1}} \quad (3)$$

The field distribution in the space is a sinusoidal function based on the space angle. With the changing of time $t$, the space field will rotate with time. Because it is a synchronous motor, there is no spatial displacement between the rotor and stator.

For recent HTS research work [5]-[7], [10], [11], racetrack HTS coils are widely employed as the field winding. Therefore this research mainly focuses on coils with a racetrack geometry. Although the current distribution in the field winding can be simply described by (2), which is applied in [13], a new electromagnetic model is still required to be developed due to HTS coil's unique geometrical structure. In this new model, the combination of a conventional armature winding and racetrack HTS coils is achieved. For the 2-D analysis, the HTS coil is considered to be infinitely long and the current in the coil flows only in and out of the page. Fig.2 shows a 2-pole motor current distribution for the HTS coils.

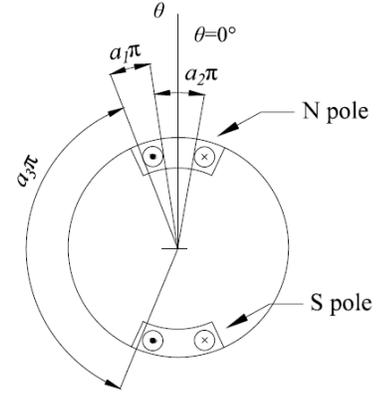

Fig. 2. Current distribution for the HTS coils in the space

In the $p$ pole-pair motor, every pole-pair occupies electrical angles around $2\pi$. For one pole-pair, one side of the racetrack coil occupies the electrical angle $a_1\pi$. The distance between one side of the coil and the other side of the coil is $a_2\pi$. The distance between one side of the coil and the side of the other coil is $a_3\pi$. This is shown diagrammatically in Fig. 2. Because the sum of all the electrical angles in one pole-pair is $2\pi$, we can obtain the following relationship:

$$2a_1 + a_2 + a_3 = 1 \quad (4)$$

Considering a current sheet $K_F(\theta)$ for the HTS coils (field winding), the calculation can be completed in terms of the number of turns in series in the HTS coil ($N_r$), the current in the HTS coil ($I_r$), the width of the coil ($w_r$), the thickness of the coil ($h_r$). Since the coil is wound by thin coated conductor, whose thickness is much smaller than the motor diameter, the magnitude of the current sheet $K_F$ in the coil can be considered as constant and can be shown to be

$$K_F = \frac{N_r I_r}{w_r h_r} h_r = \frac{N_r I_r}{w_r} \quad (5)$$

Set the space degree $\theta = 0$ from the longitudinal axis, which is the same as that of armature winding. The current sheet distribution of the racetrack HTS coils is shown in Fig. 3.

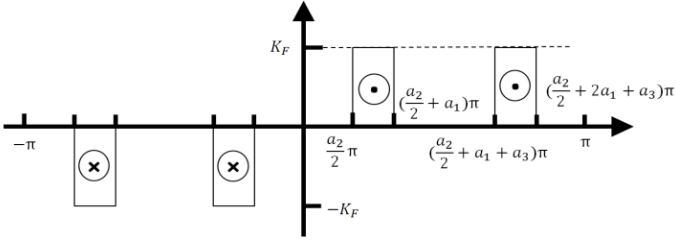

Fig. 3. The current sheet distribution of the racetrack HTS coils

For the $p$ pole-pair synchronous motor, applying Fourier analysis, the current sheet for HTS field winding can be expressed as:

$$K_F(\theta) = \sum_{n=2Z+1} K_{Fn} \sin(np\theta), z = 0,1,2... \quad (6)$$

$$K_{Fn} = \frac{8K_F}{n\pi} \sin(\frac{n(a_1+a_2)}{2}\pi) \sin(\frac{na_1}{2}\pi) \quad (7)$$

### B. Synchronous Motor Field Analysis

For this simplified 2-D model, the magnetic vector potential in Fig. 1, comprising only the component $A_z$, satisfies Laplace's equation in the regions of no current. Therefore, in cylindrical coordinates, for all regions in Fig. 1, except for the location of the armature and field windings, we have:

$$\frac{\partial^2 A_z}{\partial r^2} + \frac{1}{r}\frac{\partial A_z}{\partial r} + \frac{1}{r^2}\frac{\partial^2 A_z}{\partial \theta^2} = 0 \quad (8)$$

Considering the current sheets in both windings are given as $\sum_{n=2Z+1} K_{Fn} \sin(np\theta)$ and $\sum_{n=2mz+1} K_{An} \sin(np\theta), z = 0,1,2...$

The radial and tangential flux densities, $B_r$ and $B_\theta$ and the radial and tangential magnetic field components, $H_r$ and $H_\theta$ can be calculated as

$$B_r = \frac{1}{r}\frac{\partial A_z}{\partial \theta}, B_\theta = -\frac{\partial A_z}{\partial r}, B_r = \mu_0 H_r, B_\theta = \mu_0 H_\theta \quad (9)$$

For the boundary conditions, we have:

$A_z = 0$, when $r = 0$ and $r = \infty$,
$B_r$ continuous at $r_b, r_{s1}, r_{s2}$,
$H_\theta$ continuous at $r_{s2}$,
$H_\theta(r_{s1}^+) - H_\theta(r_{s1}^-) = \sum_{n=2mz+1} K_{An} \sin(np\theta), z = 0,1,2...$,
$H_\theta(r_b^+) - H_\theta(r_b^-) = \sum_{n=2Z+1} K_{Fn} \sin(np\theta), z = 0,1,2...$.

From (8), (9) and the conditions above, the flux densities are:

(i) Region $0 \leq r \leq r_b$ (rotor, see Fig.1)

$$B_r = \frac{\mu_0}{2}[\sum_{n=2z+1} K_{Fn}(\frac{r}{r_b})^{np-1}\cos(np\theta) + \sum_{n=2mz+1} K_{An}(\frac{r}{r_b})^{np-1}\cos(np\theta)] \quad (10)$$

$$B_\theta = \frac{\mu_0}{2}[\sum_{n=2z+1} K_{Fn}(\frac{r}{r_b})^{np-1}(-\sin(np\theta)) + \sum_{n=2mz+1} K_{An}(\frac{r}{r_{s1}})^{np-1}(-\sin(np\theta))] \quad (11)$$

(ii) Region $r_b \leq r \leq r_{s1}$ (air gap, see Fig.1)

$$B_r = \frac{\mu_0}{2}[\sum_{n=2z+1} K_{Fn}(\frac{r_b}{r})^{np+1}\cos(np\theta) + \sum_{n=2mz+1} K_{An}(\frac{r}{r_{s1}})^{np-1}\cos(np\theta)] \quad (12)$$

$$B_\theta = \frac{\mu_0}{2}[\sum_{n=2z+1} -K_{Fn}(\frac{r_b}{r})^{np+1}(-\sin(np\theta)) + \sum_{n=2mz+1} K_{An}(\frac{r}{r_{s1}})^{np-1}(-\sin(np\theta))] \quad (13)$$

(iii) Region $r > r_{s1}$ (stator, see Fig.1)

$$B_r = \frac{\mu_0}{2}[\sum_{n=2z+1} K_{Fn}(\frac{r_b}{r})^{np+1}\cos(np\theta) + \sum_{n=2mz+1} K_{An}(\frac{r_{s1}}{r})^{np+1}\cos(np\theta)] \quad (14)$$

$$B_\theta = \frac{\mu_0}{2}[\sum_{n=2z+1} -K_{Fn}(\frac{r_b}{r})^{np+1}(-\sin(np\theta)) + \sum_{n=2mz+1} -K_{An}(\frac{r_{s1}}{r})^{np+1}(-\sin(np\theta))] \quad (15)$$

For (10)-(15), $z = 0,1,2...$

### C. Synchronous Motor Reactance

The synchronous inductance in the model is related to the conventional armature winding. For a conventional winding, the fundamental harmonic winding factor $k_{w1}$ can be used to calculate the inductance. Setting $K_{Fn} = 0$, n = 1 in (10)-(15) and then evaluating the flux linkages separately, the self-inductance and mutual inductance can be calculated. Summing the self-inductance and mutual inductance, the reactance can be obtained:

$$X_s = \frac{m}{2}\omega\mu_0 \frac{\pi}{p}(\frac{N_s k_{w1}}{\pi})^2 \ \Omega \ \text{m}^{-1} \quad (16)$$

## IV. MODEL VERIFICATION

For an HTS machine, numerical analysis is one of the most effective methods to verify the analytical solution. Commercial software Comsol Multiphysics 4.3a is used to carry out the finite element analysis. The *H*-formulation is applied to construct the model, which is shown in Fig. 4. The *H*-formulation has been applied variously to solve high temperature superconductor problems for over a decade [14]–[21]. The method has the potential for better computational speed and convergence properties, and it can be more practical for integrating boundary conditions [22], [23], compared with the *A-V* [24]-[27] and *T-Ω* [28]-[30] formulations.

Considering that the size of the HTS coil is small compared with the motor's size, a bulk model has been used to replace the layers of coated conductor in the HTS coils. Therefore, the integration of the current density in each bulk element is equal to the applied current $I_r$ in the tape multiplied by the number of turns $N_r$. In order to carry out a more practical model analysis, SuperPower's SCS4050 wire [31] is chosen as an example, with a critical current of 80 A (77 K, self-field), width 4 mm, and thickness 1 μm. For a MW-class motor design, assuming the stator current is 145 A and the terminal voltage is 3000 V, the model's parameters are show in Table I. With the extension of previous examples [12], [16], [20], [23], the 2-D, and 3-D machine models are designed for comparison with the analytical solution, and the field distributions are shown in Fig. 4. The motor's field in all regions satisfies Maxwell's equations. For the simplified air-cored superconducting machine, there are only three materials: copper in the armature winding, and air and superconducting material parts in the field winding.

In order to maintain consistency with Comsol Multiphysics 4.3a, $\theta = 0$ has been set along the horizontal axis (previously the vertical axis in Fig. 2). Based on (5)-(7) and (10)-(15), Matlab R2012a has been used to complete the analytical solution using the parameters in Table I.

TABLE I
Motor Parameters for Operating Temperature of 65 K

| Parameters | Symbol | Value |
|---|---|---|
| Pole-pairs | $p$ | 3 |
| Phase number | $m$ | 3 |
| Rotor radius | $r_b$ | 0.2214 m |
| Field coil turns | $N_r$ | 1500 |
| Coil thickness | $h_r$ | 30 mm |
| Coil width (one side) | $w_r$ | 30 mm |
| Distance between coil sides | $W_c$ | 110 mm |
| Field winding critical current (77 K, self-field) [manufacturer supplied] | $I_c$ | 80 A |
| Field winding critical current density (77 K, self-field) [manufacturer supplied] | $J_{c0}$ | $2\times10^{10}$ Am$^{-2}$ |
| Operating temperature | $T_{op}$ | 65 K |
| Field winding current (65 K) | $I_r$ | 60 A |
| Field winding current density (65 K) | $J_r$ | $1.5\times10^{10}$ Am$^{-2}$ |
| n-value | $n$ | 21 |
| Armature winding position radius | $r_{arm}$ | 250 mm |
| Motor length | $L$ | 600 mm |
| Stator terminal voltage | $V_s$ | 3000 V |
| Stator current | $I_s$ | 145 A |
| Rated power output | $P_{rated}$ | 1.25 MW |

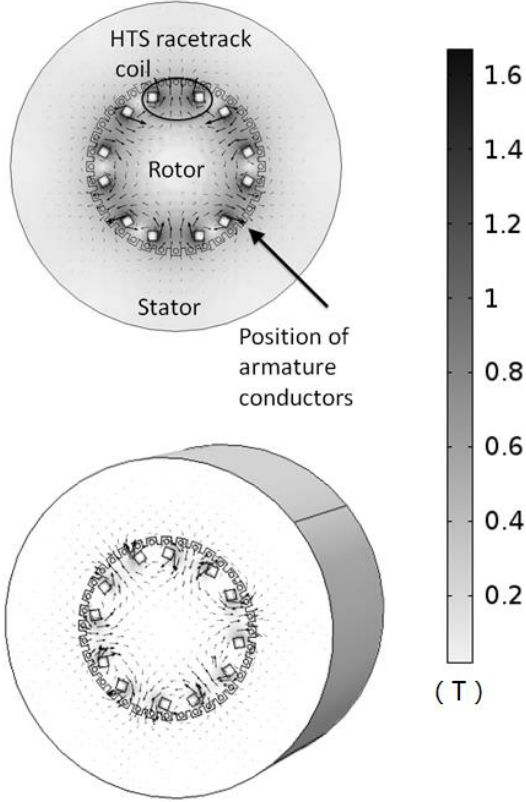

Fig. 4  2-D, 3-D geometry and magnetic field distribution using Comsol Multiphysics 4.3a

The apparent output power for an HTS electrical machine of this type is given by (17) in [13],

$$S = \frac{\pi^2}{\sqrt{2}} k_w \hat{B} A_s L D^2 n_r \quad (17)$$

Where $k_w$ is the harmonic winding factor, $\hat{B}$ is the peak value of the fundamental component of flux density at the armature winding, $A_s$ is the electrical loading of the armature, $L$ is the length of the motor, and $D$ is diameter of the motor and $n_r$ is the speed of the rotor.

Therefore, the magnetic field at the armature winding has a significant influence on performance of the motor. A comparison of the results for the magnetic field at the armature winding has been used to validate the model, which can be seen in Fig. 5. Equation (18) is used to calculate the magnitude of the magnetic field at the armature winding.

$$B_{norm} = \sqrt{B_r^2 + B_\theta^2} \quad (18)$$

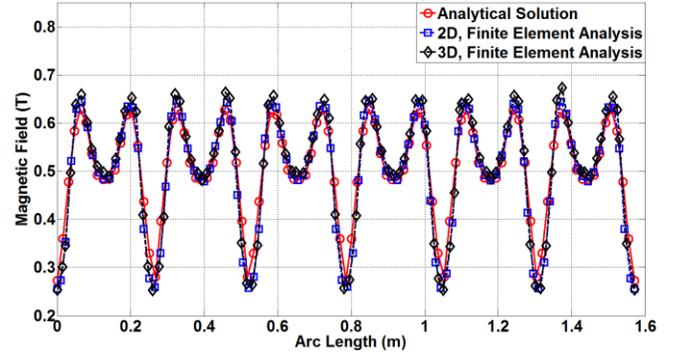

Fig. 5  Comparison of magnetic field distributions from analytical solution and 2-D and 3-D finite element analysis

It can be seen in Fig. 5 that the magnetic field distributions in the armature winding are consistent, comparing the analytical Solution, 2-D and 3-D finite element analysis.

In order to compare these further, the average $B_{norm}$ at different radii $r$ for the finite element analysis and the analytical solution are shown in Fig. 6, which again shows consistency among these three methods. In this figure, $r_f$ refers to the mean radius of the field winding.

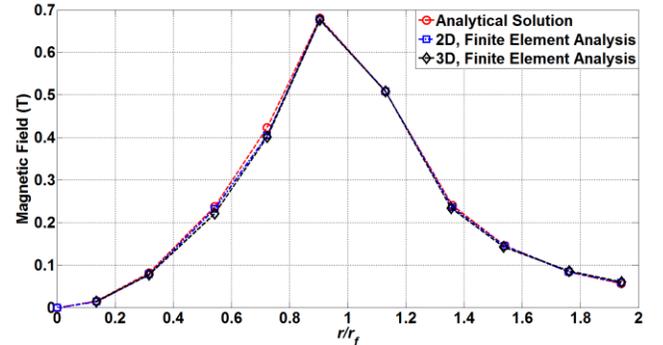

Fig. 6  Flux density with $r/r_f$

Therefore, the theoretical model has been validated successfully by numerical analysis. At the same time, the extended FEM model allows a complete integrated model, including both the electromagnetic properties of the superconductor and motor, to be developed.

## V. ADVANTAGES OF THE THEORETICAL MODEL

In fact, the analytical solution of the model is more convenient to apply than the finite element model. Firstly, the calculation of the analytical solution is much faster than that for the finite element model. Usually, the calculation of the analytical solution in Matlab will take no more than a second. For the finite element analysis, the 2-D and 3-D calculations for this model can take hours or even days, if the magnetic field distribution at a specific position is needed. Secondly, for the same model with different parameters, the theoretical model can be adjusted easily. However, for the finite element models, the geometry, mesh, boundary conditions need to be adjusted each time, which can be complicated. Thirdly, the theoretical model can help us to study the physical nature of the formulae, such as the relationship between the magnetic field and geometrical parameters. It is very useful for parametric optimization.

## VI. POTENTIAL APPLICATION OF THE THEORETICAL MODEL

### A. HTS Coil Geometry Optimization

The theoretical model can help greatly with parametric optimization. Based on (3), (5), (7), (14)-(15) and the parameters in Table I, it can be found that $K_{Fn}$ is the key factor in determining the magnetic field optimization at the armature winding in the HTS machine, and $K_{Fn}$ is related to $a_1$, $a_2$. The maximum magnetic field generated by $K_{An}$ at the armature winding is calculated to be very small (around 0.0015 T) and, therefore, optimization of $K_{An}$ is not considered here. In order to comprehensively analyze the magnetic field waveform, Fig. 7 shows the distribution of field $B_r$ at the armature winding. From this figure, the magnetic field is not a pure sinusoid, but includes a number of higher order harmonics.

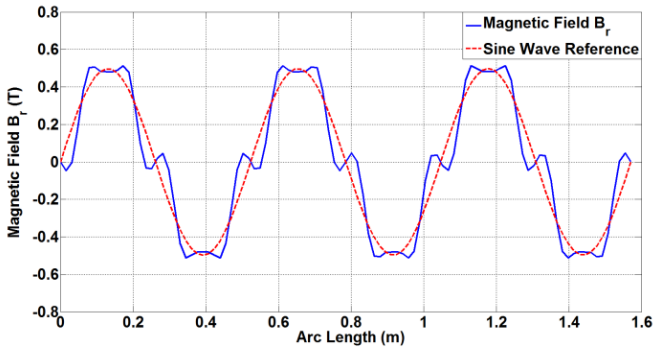

Fig. 7 Magnetic field $B_r$ at the armature winding before optimization

Ideally, the magnetic field $B_r$ at armature winding is purely sinusoidal. This means that $K_{F1}$, the fundamental harmonic of the HTS coil current, is so large that other $K_{Fn}$ (n > 1) can be ignored. These other $K_{Fn}$ can increase the harmonics in the magnetic field at armature winding, which is shown in (10)-(15). A motor's torque, $T$, is proportional to the magnetic field $B_r$ at the armature winding [13]. The presence of other harmonics at the armature winding will cause the motor torque to fluctuate, especially for high-speed rotation. Therefore, it is important to make sure that the fundamental component $K_{F1}$ is as large as possible. The harmonic distribution is shown in Fig. 8 for the waveform shown in Fig. 7.

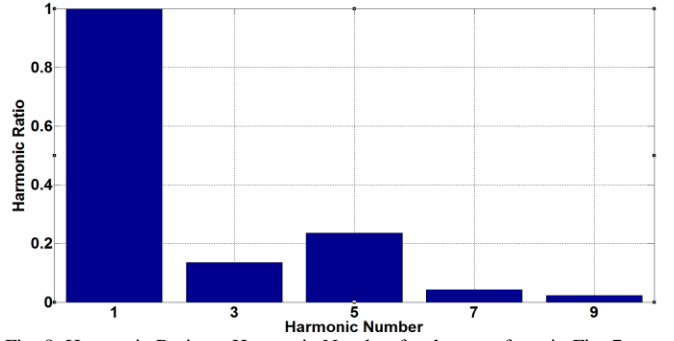

Fig. 8 Harmonic Ratio vs Harmonic Number for the waveform in Fig. 7

By harmonic analysis, the fundamental harmonic occupies almost 70% of the whole waveform, which can be improved further. The ratio of the fundamental harmonic is set as 1 in Fig. 8. From Fig. 8, the third and fifth harmonics are quite substantial, (about 0.13 and 0.23, respectively), which can also be improved.

Based on (4)-(7) and (14)-(15), optimization can be carried out as follows. From (4), the conditions are given: $0 < a_1 < 0.5$, $0 < a_2 < 1$, $2a_1+a_2 < 1$.

The SuperPower tape (SCS4050) is chosen for the HTS coils, whose specification states that the critical bend diameter in tension and in compression is 11 mm [31]. Therefore the distance between both sides of any HTS coil should be larger than $d_{coil}$ = 11 mm. Considering a rotor radius of $r$ = 0.2214 m, with the number of pole-pairs $p$ = 3, in the example, another condition is given:

$$a_2 \geq \frac{d_{coil}}{\frac{\pi r}{p}} = 0.0475 \quad (19)$$

Combining (5), (7), and the conditions above and setting the magnitude of $B_r$ = 0.5 T based on (14), which is equal to the value in Fig. 7, $K_{F1}$ is dominant and $K_{F3}$ reduces significantly when $a_1$ = 0.1246 (i.e., the coil width is 28.4 mm), $a_2$=0.5421 (i.e., the distance between the coil sides is 123.6 mm). Therefore, the coil width should become smaller, compared with 30 mm before optimization (see Table I). The distance between both sides of the coil should be larger, from 110 mm in Table I to 123.6 mm here.

After optimization, the flux density at the armature winding is shown in Fig. 9.

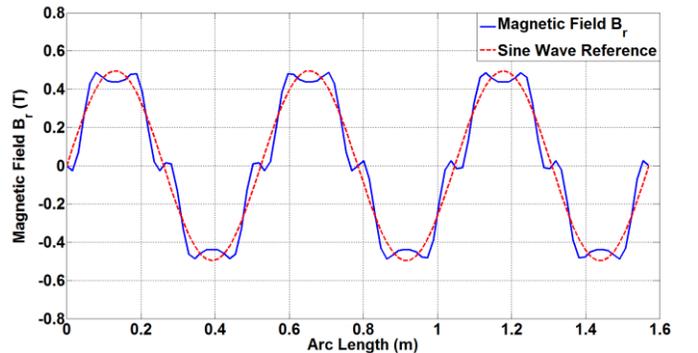

Fig. 9 Magnetic field $B_r$ at the armature winding after optimization

Comparing Figs. 7 and 9, it may be difficult to visualize the results of optimization; therefore, the harmonics of the magnetic field are shown in Fig. 10.

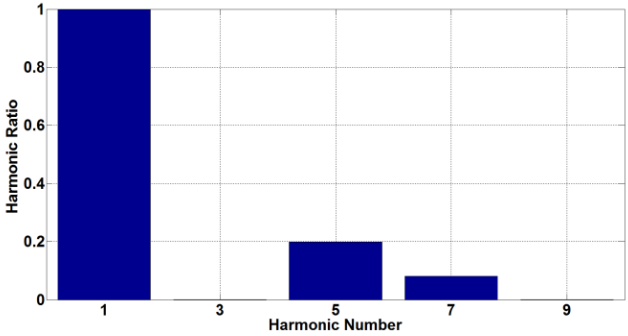
Fig. 10 Harmonic Ratio vs Harmonic Number after optimization

By harmonic analysis, the percentage of the fundamental harmonic waveform (in terms of the whole waveform) has increased by about 12% (from 70% to 82%). The ratio of the fundamental harmonic is set as 1 in Fig. 10. Comparing Figs. 8 and 10, the third harmonic has almost been removed and the fifth harmonic has also decreased, from 0.23 in Fig. 8 to 0.2 in Fig. 10. Although the 7th harmonic waveform has increased somewhat, the magnitude is still quite small. Total harmonic distortion (THD) of the field is given by (20) to analyze the quality of the field in electric machine.

$$\text{THD} = \frac{\sqrt{\sum_{n=2}^{\infty} B_{rn}^2}}{B_{r1}} \quad (20)$$

In (20), $B_{rn}$ represents the $n^{th}$ harmonic of the perpendicular magnetic field and $B_{r1}$ represents the fundamental harmonic. By optimization, THD decreases from 0.276 to 0.213, and more sinusoid magnetic field is obtained. Therefore, the optimization is considered successful.

*B. Kim-like Model for Superconducting Properties*

Although the waveform has been optimized successfully, the current 60 A (65 K) in the tape should be judged whether it is within a safe operating range. By Matlab, the maximum perpendicular magnetic field seen by the HTS tape (maximum parallel magnetic field in the HTS coils) is close to 1.5 T. In this analysis, it is assumed that the critical current decreases most significantly due to magnetic fields perpendicular to the tape face, which allows the application of a Kim-like model [12], shown in (21), although this is not necessarily true for all 2G YBCO tapes [32], [33].

$$J_c(B_{r1}, T) = J_{c0}(T) \times \frac{1}{1 + \frac{B_{rL}}{B_0}} \quad (21)$$

where $B_{r1}$ is the magnetic field perpendicular to the superconductor tape surface, $J_c(B_{r1}, T)$ is the critical current density at specific magnetic field $B_{r1}$ and specific temperature $T$, $J_{c0}(T)$ is the critical current density in self-field at specific temperature $T$. $B_0$ is constant, which differs between superconducting materials. The data in Fig. 11 comes from [34], which shows the perpendicular field dependence of critical current at 65 K. Based on this data, curve fitting is completed based on (21), which is also shown in Fig. 11. $B_0$ has been calculated, which is 0.8978 and $J_{c0}$ (0 T, 65 K) = 2.481 × $J_{c0}$ (0 T, 77 K) = 5 × $10^{10}$ Am$^{-2}$.

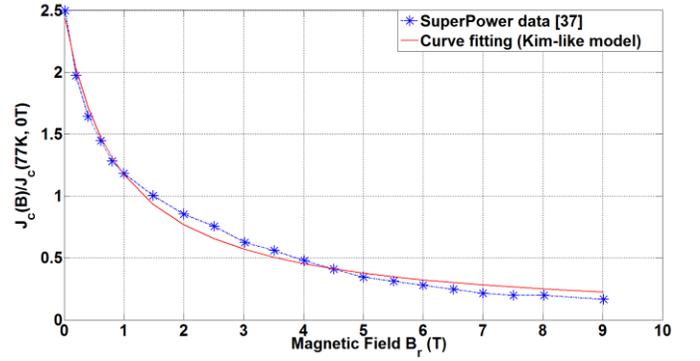
Fig. 11 Practical data in the superpower and curve fitting

For a perpendicular magnetic field at the HTS coil closed to 1.5 T and based on (21), $J_{c0}$ (1.5 T, 65 K) = 1.9 × $10^{10}$ Am$^{-2}$ > 1.5× $10^{10}$ Am$^{-2}$ (practical operating current in the HTS coil) and therefore the practical current is about 78% of the maximum allowable current at 65 K. Therefore, a current of 60 A in the 4 mm HTS tape is considered safe to be used in this motor design, for an operating temperature of 65 K.

*C. Further Optimization of Magnetic Field*

The distribution of the coils has been optimized. However, the magnitude of $B_r$ = 0.5 T at the armature winding (see Fig. 9) would not exploit the advantages of using HTS coils. Therefore, here we increase the magnitude to more than 1 T by setting operating current density as 2.875 x $10^{10}$ Am$^{-2}$. Carrying out the same process (A in Section VI), $a_1$ = 0.32 (coil width 75.9 mm), $a_2$ = 0.33 (distance between coil sides is 76 mm).

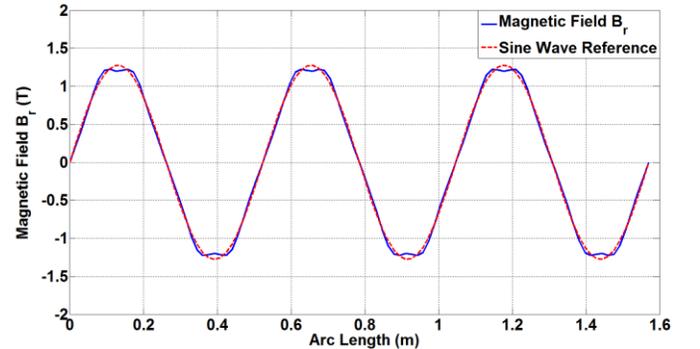
Fig. 12 Magnetic fields $B_r$ at armature winding

From Fig. 12, the magnetic field $B_r$ at armature winding is almost a pure sinusoid (see reference waveform) and its magnitude is about 1.2 T. The magnetic field $B_{r1}$ at the HTS coil (field winding) is close to 2.5 T. By calculation using (21), the critical current at 65 K and 2.5 T is 1.32 x $10^{10}$ Am$^{-2}$ < 2.875×$10^{10}$ Am$^{-2}$ (the actual operating current in the HTS coil). Therefore, a lower operating temperature is required. Based on the data in [34], 40 K has been chosen and the new optimized motor parameters are shown in Table II.

TABLE II
Motor Parameters with Operation Temperature 40 K

| Parameters | Symbol | Value |
|---|---|---|
| Pole-pairs | $p$ | 3 |
| Phase number | $m$ | 3 |
| Rotor radius | $r_b$ | 0.2214 m |
| Field coil turns | $N_r$ | 1980 |
| Coil thickness | $h_r$ | 30 mm |
| Coil width (one side) | $w_r$ | 75.9 mm |
| Distance between coil sides | $W_c$ | 76 mm |
| Field winding critical current (77 K, self-field) [manufacturer supplied] | $I_c$ | 80 A |
| Field winding critical current density (77 K, self-field) [manufacturer supplied] | $J_{c0}$ | $2 \times 10^{10}$ Am$^{-2}$ |
| Operating temperature | $T_{op}$ | 65 K |
| Field winding current (40 K) | $I_r$ | 115 A (80% of critical current) |
| Field winding current density (40 K) | $J_r$ | $2.875 \times 10^{10}$ Am$^{-2}$ |
| n-value | $n$ | 21 |
| Armature winding position radius | $r_{arm}$ | 250 mm |
| Motor length | $L$ | 600 mm |
| Stator terminal voltage | $V_s$ | 7200 V |
| Stator current | $I_s$ | 145 A |
| Rated power output | $P_{rated}$ | 3 MW |

## VII. CONCLUSION

The paper proposes a theoretical model of an air-cored HTS synchronous motor with HTS racetrack coils as the field winding. To optimize the design parameters and performance of such a machine, a basic physical model of an air-cored HTS synchronous motor was presented, as well as 2-D, and 3-D numerical FEM models to verify the analytical solution in terms of the magnetic field produced by both the conventional armature winding and the HTS field winding. The analytical solution of the theoretical model can be used in a much easier way than FEM models, and has numerous advantages and potential applications. Furthermore, the finite element model developed in this paper allows a complete integrated model, including both the electromagnetic properties of the superconductor and motor, to be developed. The analytical model is utilized to study the influence of the geometry of the HTS coils on the magnetic field at the armature winding, and geometrical parameter optimization is carried out to obtain a more sinusoidal magnetic field at the armature winding, which improves motor performance.

## VIII. ACKNOWLEDGEMENT

The authors would like to thank Prof. Archie Campbell, and Drs Patrick Palmer and Tim Coombs for their helpful discussion on the original version of this document.

## IX. REFERENCES

[1] S. S. Kalsi, K. Weeber, H. Takesue, C. Lewis, H-W. Neumueller, and R.D. Blaugher, "Development status of rotating machines employing superconducting field windings," *Proc. IEEE,* vol. 92, pp. 1688-1704, Oct. 2004.
[2] T. Nakamura, Y. Yamada, H. Nishio, K. Kajikawa, M. Sugano, N. Amemiya, T. Wakuda, M. Takahashi, and M. Okada, "Development and fundamental study on a superconducting induction/synchronous motor incorporated with MgB2 cage windings," *Supercond. Sci. Technol.*, vol. 25, pp. 014004, Dec. 2011.
[3] Y. Terao, M. Sekino, and H. Ohsaki, "Electromagnetic design of 10 MW class fully superconducting wind turbine generators," *IEEE Trans. Appl. Supercond.*, vol. 22, pp. 5201904, Jun. 2012.
[4] D. Zhou, M. Izumi, M. Miki, B. Felder, T. Ida, and M. Kitano, "An overview of rotating machine systems with high-temperature bulk superconductors," *Supercond. Sci. Technol.*, vol. 25, pp. 103001, Aug. 2012.
[5] G. Snitchler, B. Gamble, and S. S. Kalsi, "The Performance of a 5MW High Temperature Superconductor Ship Propulsion Motor," *IEEE Trans. Appl. Supercond.* vol. 15, pp. 2206-2209, Jun. 2005.
[6] J.M. Fogarty, "Development of a 100 MVA high temperature superconducting generator," *Power Engineering Society General Meeting*, pp. 2065-2067, Jun. 2004.
[7] M. Frank, J. Frauenhofer, B. Gromoll, P. van Haßelt, W. Nick, G. Nerowski, H-W. Neumüller, H-U. Häfner, and G. Thummes, "Thermosyphon cooling system for the siemens 400kW HTS synchronous machine," *AIP Conf. Proc.* vol. 710, pp. 859, Sep. 2003.
[8] H-W. Neumüller, W. Nick, B. Wacker, M. Frank, G. Nerowski, J. Frauenhofer, W. Rzadki, and R. Hartig, "Advances in and prospects for development of high-temperature superconductor rotating machines at Siemens," *Supercond. Sci. Technol.,* vol. 19, S114, Mar. 2006.
[9] W. Nick, G. Nerowski, H-W. Neumüller, M. Frank, P. van Hasselt, J. Frauenhofer, and F. Steinmeyer, "380 kW synchronous machine with HTS rotor windings – development at Siemens and first test results," *Physica C,* vol. 372, pp. 1506-1512, Aug. 2002.
[10] W. O. S. Bailey, M. Al-Mosawi, Y. Yang, K. Goddard, and C. Beduz, "The design of a lightweight HTS synchronous generator cooled by sub-cooled liquid nitrogen," *IEEE Trans. Appl. Supercond.*, vol. 19, pp. 1674-1677, Jun. 2009.
[11] S. Fukui, T. Kawai, M. Takahashi, J. Ogawa, T. Oka, T. Sato, and O. Tsukamoto, "Numerical Study of Optimization Design of High Temperature Superconducting Field Winding in 20MW Synchronous Motor for Ship Propulsion," *IEEE Trans. Appl. Supercond.*, vol. 22, pp. 5200504, Jun. 2012.
[12] M. D. Ainslie, V. Zermeno, Z. Hong, W. Yuan, T. J. Flack and T. A. Coombs, "An improved FEM model for computing transport AC loss in coils made of RABiTS YBCO coated conductors for electric machines," *Supercond. Sci. Technol.,* vol. 24, pp. 075001, Apr. 2011.
[13] J. R. Bumby, *Superconducting Rotating Electrical Machines*. Oxford: Clarendon, 1983, pp. 71-74, 106-109.
[14] R. Pecher, M. D. McCulloch, S. J. Chapman, L. Prigozhin, and C. M. Elliott, "3D-modelling of bulk type-II superconducting using unconstrained H-formulation," *Proceedings of the 6th EUCAS*, p. 1-11, 2003.
[15] K. Kajikawa, T. Hayashi, R. Yoshida, M. Iwakuma, and K. Funaki, "Numerical evaluation of AC losses in HTS wires in 2D FEM formulated by self-magnetic field," *IEEE Trans. Appl. Supercond.*, vol. 13, pp. 3630-3633, Jun. 2003.
[16] Z. Hong, A. M. Campbell, and T. A. Coombs, "Numerical solution of critical state in superconductivity by finite element software," *Supercond. Sci. Technol.* vol. 19, pp. 1246, Dec. 2006.
[17] R. Brambilla, F. Grilli, and L. Martini, "Development of an edge-element model for AC loss computation of high-temperature superconductors," *Supercond. Sci. Technol.*, vol. 20, pp. 16-24, Jan. 2007.
[18] F. Sirois, M. Dione, F. Roy, F. Grilli, and B. Dutoit, "Evaluation of two commercial finite element packages for calculating AC losses in 2-D high temperature superconducting strips," *J. Phys.: Conf. Ser.,* vol. 97, pp. 012030, 2008.
[19] M. D. Ainslie, T. J. Flack, and A. M. Campbell, "Calculating transport AC loss in stacks of high temperature superconductor coated conductors with magnetic substrates using FEM," *Physica C*, vol. 472, pp. 50-56, Jan. 2012.
[20] M. D. Ainslie, W. Yuan, and T. J. Flack, "Numerical analysis of AC loss reduction in HTS Superconducting Coils Using magnetic Materials to Divert Flux," *IEEE Trans. Appl. Supercond.,* vol. 23, pp. 4700104, Jun. 2013.
[21] V. Zermeno, F. Grilli, and F. Sirois, "A full 3D time-dependent electromagnetic model for Roebel cables," *Supercond. Sci. Technol*, vol. 26, pp. 052001, May. 2013.

[22] M. D. Ainslie, T. J. Flack, Z. Hong, and T. A. Coombs, "Comparison of first- and second-order 2D finite element models for calculating AC loss in high temperature superconductor coated conductors," *Int. J. Comput. Math. Electr. Electron. Eng.,* vol. 30, pp. 762-774, 2011.
[23] M. Zhang and T. A. Coombs, "3D modelling of high-Tc superconductors by finite element software," *Supercond. Sci. Technol.,* vol. 25, pp. 015009, Jan. 2012.
[24] L. Prigozhin, "Analysis of critical-state problems in type-II superconductivity," *IEEE Trans. Appl. Supercond.* vol. 7, pp. 3866, Dec. 1997.
[25] G. Barnes, M. McCulloch and D. Dew-Hughes, "Computer modelling of type II superconductors in applications", *Supercond. Sci. Technol.* 12, pp. 518, Aug. 1999.
[26] S. Stavrev, F. Grilli, B. Dutoit, N. Nibbio, E. Vinot, I. Klutsch, G. Meunier, P. Tixador, Y. Yang, and E. Martinez, "Comparison of numerical methods for modeling of superconductors," *IEEE Trans. Magn.*, vol. 38, pp. 849, Mar. 2002.
[27] A. M. Campbell, "A direct method for obtaining the critical state in two and three dimensions," *Supercond. Sci. Technol*. 22, pp. 034005, Mar. 2009.
[28] N. Amemiya, S. Murasawa, N. Banno, and K. Miyamoto, "Numerical modelings of superconducting wires for AC loss calculations," *Physica C,* vol. 310, pp. 16, Dec. 1998.
[29] N. Amemiya, K. Miyamoto, S. Murasawa, H. Mukai, and K. Ohmatsu, "Finite element analysis of AC loss in non-twisted Bi-2223 tape carrying AC transport current and/or exposed to DC or AC external magnetic field," *Physica C*, vol. 310, pp. 30–35, Dec. 1998.
[30] G. Meunier, Y. Floch, and C. Guerin, "A nonlinear circuit coupled t-t$_0$-φ formulation for solid conductors," *IEEE Trans. Magn*., vol.39, pp. 1729, May 2003.
[31] SuperPower Inc. http://www.superpower-inc.com
[32] F. Gomory, J. Souc, E. Pardo, E. Seiler, M. Soloviov, L. Frolek, M. Skarba, P. Konopka, M. Pekarcikova, and J. Janovec, "AC loss in pancake coil made from 12 mm wide REBCO tape," *IEEE Trans. Appl. Supercond.,* vol. 23, Jun. 2013.
[33] E. Pardo, M. Vojenciak, F. Gomory, and J. Souc, "Low-magnetic-field dependence and anisotropy of the critical current density in coated conductors," *Supercond. Sci. Technol*., vol. 24, pp. 065007, Jun. 2011.
[34] V. Selvamanickam, Y. Yao, Y. Chen, T. Shi, Y. Liu, N. D. Khatri, C. Lei, E. Galstyan, and G. Majkic, "The low-temperature, high-magnetic-field critical current characteristics of Zr-added (Gd,Y)Ba2Cu3Ox superconducting tapes," *Supercond. Sci. Technol*., vol. 25, pp. 125013, Dec. 2012.

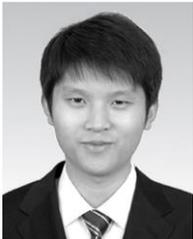

**Di Hu** was born in Wuhan, China, in 1990. He received the B.Eng degree in electronic and electrical engineering from both Huazhong University of Science and Technology and the University of Birmingham in 2012, and is currently working toward the Ph.D. degree in engineering at the University of Cambridge, Cambridge, UK

His areas of interests include the theoretical analysis and simulation via the finite element method of high temperature superconductor (HTS) machines, and research on the electromagnetic properties of HTS tape and bulk materials for electric machines.

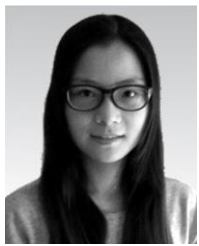

**Jin Zou** was born in Henan Province, China, in 1989. She received the B.Eng. from both the University of Birmingham, UK and Central South University, China, in 2012. She is currently a Ph.D. student at the University of Cambridge. Her main research interests include the simulation and control of high temperature superconducting machines.

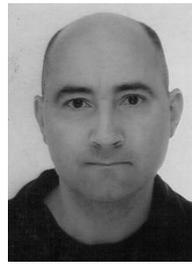

**Tim Flack** was awarded his B.Sc in Electrical Engineering from Imperial College, London, in 1986 and obtained his PhD from the same institution in 1990. Since then he has worked as a post-doc, and then as a University Lecturer at the Cambridge University Engineering Department.

His research interests include: the development of 2-D and 3-D finite-element codes for the modelling of electrical machines and also nano and micromagnetic simulations; electromagnetically-geared motors and generators; applied superconductivity and its modelling using the finite-element method; development of the brushless doubly-fed generator for wind power.

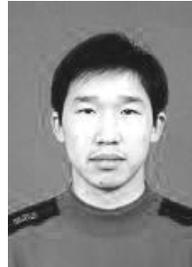

**Xiaozhuo Xu** was born in China in 1980, and received the B.S. and M.S. degrees in electrical engineering and automation, motor and electrical from the School of Electrical Engineering and Automation, Henan Polytechnic University, China, in 2003 and 2006, respectively. His research interests are the analysis of physical fields for special motors, and the optimization design of linear and rotary machines.

He is currently a lecturer in the School of Electrical Engineering and Automation, Henan Polytechnic University, China.

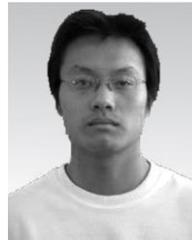

**Haichao Feng** was born in China in 1983, and received the B.S. and M.S. degrees in electrical engineering and automation, control theory and control engineering from the School of Electrical Engineering and Automation, Henan Polytechnic University, China, in 2005 and 2008, respectively. He is currently a teacher in the School of Electrical Engineering and Automation, Henan Polytechnic University, China.

Mr. Feng is a member of the Institute of Linear Electric Machines and Drives, Henan Province, China.

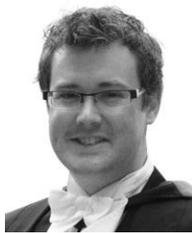

**Mark Douglas Ainslie** (M'2005) was born in Adelaide, Australia, in 1981. He received the B.E. & B.A. (Japanese) degree from the University of Adelaide, Adelaide, Australia, in 2004, the M.Eng. degree from the University of Tokyo, Tokyo, Japan, in 2008, and the Ph.D. degree from the University of Cambridge, Cambridge, UK, in 2012.

He is currently a Royal Academy of Engineering Research Fellow in the Bulk Superconductivity Group at the University of Cambridge and is a Junior Research Fellow at King's College, Cambridge. His current research interests are in applied superconductivity in electrical engineering, including superconducting electric machine design, power system protection and energy storage, electromagnetic modeling, and interactions between conventional and superconducting materials.